\documentclass[journal]{IEEEtran}

\usepackage{cite}
\usepackage{amsmath,amssymb,amsfonts}
\usepackage{algorithm}
\usepackage{subfig}
\usepackage{amsfonts}
\usepackage{amssymb}
\usepackage{gensymb}
\usepackage{multirow}
\usepackage{booktabs}
\usepackage{algorithmic}
\usepackage{textcomp}
\usepackage[normalem]{ulem}
\usepackage{xcolor}
\def\BibTeX{{\rm B\kern-.05em{\sc i\kern-.025em b}\kern-.08em
    T\kern-.1667em\lower.7ex\hbox{E}\kern-.125emX}}
\ifCLASSINFOpdf
   \usepackage[pdftex]{graphicx}
   \graphicspath{{../pdf/}{../jpeg/}}
   \DeclareGraphicsExtensions{.pdf,.jpeg,.png}
\else
   \usepackage[dvips]{graphicx}
   \graphicspath{{../eps/}}
   \DeclareGraphicsExtensions{.eps}
\fi
\usepackage{amsmath}
\usepackage{amssymb}
\usepackage{subfig}

\begin{document}

\title{Widely Distributed Radar Imaging: Unmediated ADMM Based Approach}
%
%
%

\author{Ahmed~Murtada,~Ruizhi~Hu,~Bhavani~Shankar~Mysore~Rama~Rao,~Udo~Schroeder
\thanks{This work is supported by Luxembourg National Research Fund (FNR) through the Industrial Fellowship project "RADII", ref. 15364040.} 
\thanks{A. Murtada, R. Hu, and B.S.M. Rama Rao are with the Interdisciplinary Centre for Security, Reliability and Trust (SnT), University of Luxembourg, L1885 Luxembourg City, Luxembourg (e-mail: ahmed.murtada@uni.lu; ruizhi.hu@uni.lu; bhavani.shankar@uni.lu). U. Schroeder is with IEE S.A., L-7795 Bissen, Luxembourg (email: Udo.schroeder@iee.lu).}
}

%
%

\markboth{}%
{Shell \MakeLowercase{\textit{et al.}}: Bare Demo of IEEEtran.cls for IEEE Journals}
%



\maketitle

\begin{abstract}
In this paper, we present a novel approach to reconstruct a unique image of an observed scene with widely distributed radar sensors. The problem is posed as a constrained optimization problem in which the global image which represents the aggregate view of the sensors is a decision variable. While the problem is designed to promote a sparse solution for the global image, it is constrained such that a relationship with local images that can be reconstructed using the measurements at each sensor is respected. Two problem formulations are introduced by stipulating two different establishments of that relationship. The proposed formulations are designed according to consensus ADMM (CADMM) and sharing ADMM (SADMM), and their solutions are provided accordingly as iterative algorithms. We drive the explicit variable updates for each algorithm in addition to the recommended scheme for hybrid parallel implementation on the distributed sensors and a central processing unit. Our algorithms are validated and their performance is evaluated exploiting Civilian Vehicles Dome data-set to realize different scenarios of practical relevance. Experimental results show the effectiveness of the proposed algorithms especially in cases with limited measurements.
\end{abstract}


%
\IEEEpeerreviewmaketitle

\section{Introduction}
\label{sec:intro}

Widely distributed radar systems are robust and fault-tolerant systems that provide high angular resolution and permit the exploitation of spatial diversity and occlusions avoidance \cite{haimovich2007mimo}. With applications in surveillance, assisted living, and health monitoring, radar systems with widely distributed antennas are expected to play a vital role in emerging sensing paradigms \cite{gurbuz2019radar,gennarelli2019radar}. Under this architecture, the observed targets feature an aspect dependent scattering behavior restricting the employment of conventional imaging methods. The main impediment arises due to the adoption of the isotropic point scattering model of targets, thereby preventing algorithms like back-projection (BP) to provide an adequate imaging performance \cite{moses_wide-angle_2004}.\par
The problem of radar imaging with widely distributed sensors has not received enough attention in the literature. The works \cite{lodhi_coherent_2019,mansour_sparse_2018,mansour2018radar} considered radar imaging with distributed antennas and the related issues due to ambiguity in antenna positions and clocks synchronization. In these works, model-based optimization algorithms are utilized to jointly achieve the imaging task and resolve such issues. However, in these works, an isotropic scattering model, suitable when antennas are closely spaced, is assumed; this is clearly not suitable when widely separated antennas are considered. On the other hand, in wide-angle synthetic aperture radars (WSAR), which bear a close resemblance to a widely distributed architecture, two approaches exist for imaging \cite{ash_wide-angle_2014}. The first one is based on parametric modeling that characterizes the canonical scattering behavior of scatterers \cite{potter1997attributed,liu_efficient_2019,sugavanam_interrupted_2017,yang_robust_2019}. Correspondingly, the scene image is reconstructed through joint processing of the measurements from the whole aperture exploiting the model. Nevertheless, the imaging involves a dictionary search process that is computationally cumbersome \cite{hammond2013sar}. The other approach is composite imaging \cite{hu_video-sar_2017,sanders_composite_2017,wei_wide_2018,xu_accurate_2021} in which the full aperture is divided into sub-apertures within which the point scattering model holds. Respectively, images of each sub-aperture are formed through regularized optimization exploiting specific features such as sparsity. As a final step, individual images are fused to constitute an aggregate image of the scene through simple techniques such as the generalized likelihood ratio test (GLRT). This approach does not fully exploit the information from different aspects where the final image of the scene is only a fused version of the images reconstructed with sub-aperture data.\par
While in this paper we propose a sub-aperture method, unlike composite imaging, we propose to solve the problem of widely distributed radar imaging by directly reconstructing a global image that is introduced as an aggregate view of the scene. Besides, the prior information is only imposed on the global image rather than the local images of individual sensors. Concurrently, the correspondence between the local images and the global one is defined as a constraint to the optimization problem. Our approach allows for better data exploitation by including the global image as a decision variable in the optimization problem. We then provide a solution based on Alternating direction method of multipliers (ADMM) framework \cite{boyd_distributed_2011}. ADMM is a powerful distributed optimization regime suitable for systems that incorporate collection of measurements through a distributed architecture. In \cite{afonso_fast_2010} and \cite{afonso_augmented_2011}, ADMM has been introduced as a fast reconstruction method for generic imaging inverse problems. Further, in \cite{guven_augmented_2016} it is applied to reconstruct complex SAR images with enhanced features in particular, and to perform imaging with undersampled measurements in the presence of phase errors in \cite{guven_autofocused_2017}.\par
While in these works ADMM has been mainly utilized to facilitate the solution of a non-constrained optimization problem by the virtue of variable splitting, we employ its constrained formulation directly in the interest of exploiting the system architecture and implementing parallelizable image reconstruction algorithms. Accordingly, we establish two problem formulations inspired by consensus ADMM (CADMM) and sharing ADMM (SADMM). The first formulation comes as a generalization of our previous work \cite{hu_widely-distributed_2021} in which CADMM is utilized to mitigate the layover artifacts in widely distributed radar imaging by considering sub-aperture measurements from different elevations. In this work, however, we present CADMM formulation to introduce the association between sub-aperture images and the global image, and generally reconstruct the image of the scene without restriction on data viewing angles. Moreover, by stipulating the more relaxed sharing association in the constraints, we introduce the second problem formulation based on SADMM. The different association introduced by SADMM formulation enables another exploitation of the relationship between the data collected by the sub-apertures. Additionally, it provides an alternative realization of the system architecture through the ensuing unalike solution. We provide the solutions as iterative algorithms with a recommendation of a parallel implementation paradigm. Finally, Civilian Vehicles Dome data-set \cite{dungan_civilian_2010} is used to realize three experiments which comprise different practical use cases. Through them, we validate our algorithms and show the performance of CADMM and SADMM, where the latter is found to provide an enhanced imaging performance in most of the scenarios. Our proposed approach can be regarded as a general framework suitable to be implemented on various architectures including WSAR and radar systems with collocated antennas.\par
\section{Signal Model and Background}
\label{sec:SigMod}
In this section, we provide the signal model we adopted for our distributed architecture and provide background about the imaging problem formulation in the state of the art.\par
Throughout this paper, vectors are denoted by lower case bold font, while matrices are in uppercase bold. $\mathbf{I}_{L}$ is the identity matrix of size $L\times L$ and $\mathbf{1}_{N}$ is a vector of all ones of size $N \times 1$. The superscripts $\mathbf{.}^{T}$ and $\mathbf{.}^{H}$ denote respectively the transpose and the complex conjugate transpose of a vector or a matrix. On the other hand, superscripts in parenthesis denote the iteration count. The symbol $\otimes$ is used for the Kronecker product.\par
%
\begin{figure}[!htbp]
	\centering
	\includegraphics[width=3.2 in]{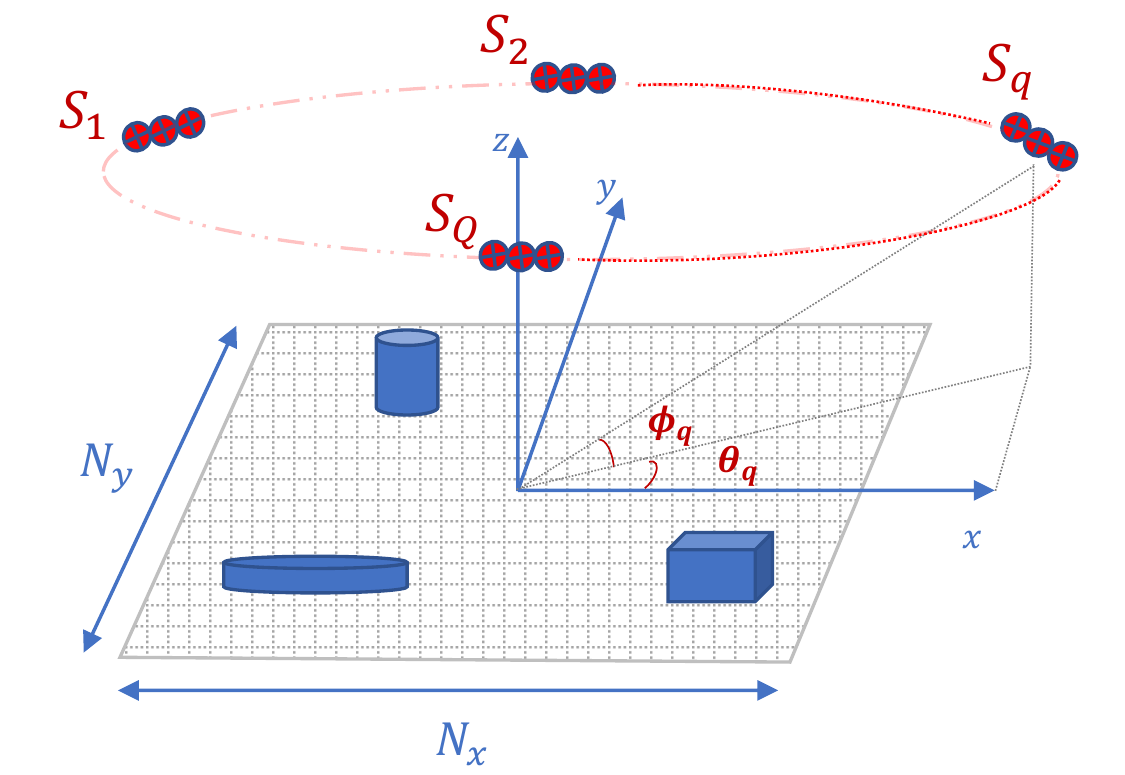}
	\caption{Geometry of Distributed Radar System}
	\label{fig:Geo}
\end{figure}
Considering the system geometry illustrated in Fig.~\ref{fig:Geo}, a group of red crossed circles constitute a cluster of antenna phase centers (APCs). The figure shows the case we consider in our paper where the $Q$ sensors would each form a single cluster are at identical elevation angles. We consider a mono-static configuration where each sensor receives the reflections due its own illumination of the scene and does not process the reflections induced by transmissions from others. Accordingly, our proposed algorithms can be applied for architectures that be formed either by a real or a synthetic aperture. At each cluster, the isotropic scattering model of the targets in the scene is assumed. This way, the problem of aspect-dependant scattering behavior can be relaxed and $Q$ local images can be formed by processing the measurements of individual clusters.\par
Since our goal is to form a reflectivity image of the scene, we adopt the 2D tomographic radar imaging framework \cite{munson1983tomographic}. Accordingly, the signal received at the $q^\text{th}$ cluster after de-chirping is 
\begin{equation}
\label{eq:radon}
{y_{q}}\left( {w,m} \right) = \iint {\tilde{x}_{q}\left({x,y} \right){e^{j \frac{4 \pi f_{w} \cos \varphi_{q}}{c}\left( {x\cos {\theta _m} + y\sin {\theta _m}} \right)}}dx \ dy},
\end{equation}
where $w = {1, \ldots, W}$ is the index of the sampled fast time frequency, $m = {1, \ldots, M}$ is the index of an APC within the cluster, $\tilde{x}_{q}\left( {x,y} \right)$ indicates the complex reflectivity coefficient of a ground target at coordinates $\left({x,y}\right)$ with respect to the $q^\text{th}$ cluster, $f_{w}$ denotes the beat linear frequency, ${\theta _m}$ is the azimuth angle of the $m^\text{th}$ element, and $\varphi_{q}$ is the elevation angle of the $q^\text{th}$ cluster.
Approximating the scene with a uniform grid of $N = N_{x} \times N_{y}$ pixels and stacking the $M$ vectors containing the frequency domain samples received by the APCs in the $q^\text{th}$ sensor, the phase history measurements can be written in a matrix form as 
\begin{equation}
\label{eq:ModelMat}
\mathbf{y}_{q}=\mathbf{A}_{q}\tilde{\mathbf{x}}_{q}+\mathbf{n}_{q}\in \mathbb{C}^{WM\times 1},
\end{equation}
where $\mathbf{A}_{q} \in \mathbb{C}^{WM \times N}$ is the system model based forward operator, $\tilde{\mathbf{x}}_{q} \in \mathbb{C}^{N\times 1}$ is the vector containing the complex scattering coefficients of the entire scene with respect to the $q^\text{th}$ cluster, and $\mathbf{n}_{q}\in \mathbb{C}^{WM \times 1}$ summarizes all errors including receiver and measurement noise as well as model imperfections.\par
Composite imaging algorithms obtain local images utilizing the signal received at each cluster and subsequently fuse them into a global image. The scene size is usually much larger than the number of measurements $N>>WM$ and the imaging task is the inverse problem of (\ref{eq:ModelMat}) which, consequently, becomes ill-posed. Compressed sensing methods are commonly used to solve this inverse problem. Particularly, local images are obtained by solving $Q$ regularized least square optimization problems for each cluster of the form
\begin{equation}
\label{eq:MBR}
\mathbf{\hat{x}}_{q}={\text{arg}}\mathop{\min }\limits_{{\tilde{\mathbf{x}}_{q}}}\left\{\left\| \mathbf{y}_{q}-\mathbf{A}_{q}\tilde{\mathbf{x}}_{q} \right\|_{2}^{2}+ h(\tilde{\mathbf{x}}_{q}) \right\},
\end{equation}
where $\mathbf{\hat{x}}_{q}$ is the estimated local image using the measurements $\mathbf{y}_{q}$ for $q=1,\ldots,Q$ and $h(\cdot)$ is a regularization function that imposes apriori information about local images. Different choices of regularization function $h(\cdot)$ exist to enhance some image features such as sparsity and smoothness, among others. when $h(\cdot)$ is a separable function (e.g. $l_1$-norm), the $Q$ problems can be represented as a single optimization problem in $Q$ variables since the least squares term is naturally separable. Explicitly, the problem can be written as
\begin{equation}
\label{eq:OptMod}
\left\{\mathbf{\hat{x}}_{1},\ldots,\mathbf{\hat{x}}_{Q} \right\}= \underset{\tilde{\mathbf{x}}_{1},\tilde{\mathbf{x}}_{2},\cdots ,\tilde{\mathbf{x}}_{Q}}{\min}\sum_{q=1}^{Q} \left\{{ \left\| \mathbf{y}_{q}-\mathbf{A}_{q}\tilde{\mathbf{x}}_{q} \right\| _{2}^{2}} + \left\| {\tilde{\mathbf{{x}}}_{q}} \right\|_1\right\}
\end{equation}
The problem in (\ref{eq:OptMod}) is an unconstrained regularized optimization problem which has been tackled through different optimization techniques in the literature. Finally, the image of the scene is obtained through a fusion step of the $Q$ reconstructed images which can be as simple as a pixel-wise maximization among the $Q$ local images.\par
As mentioned in the introduction, we alternatively reconstruct the global image of the scene by introducing its variable in the objective function and imposing the $l_1$-norm on it directly for a sparsity-driven solution. Simultaneously, the relationship between the global image and local images is defined as a constraint for our optimization problem. In the next section, based on ADMM framework, we provide two alternative problem formulations along with their solutions.
\section{ADMM Framework for Distributed Radar Imaging}
\label{sec:ADMM}
ADMM is a powerful framework that renders itself amenable for optimization problems of distributed nature. It is a suitable tool to be utilized in a distributed radar system especially when the component sensors are equipped with some computation power capabilities. Although this computation power might be limited, it can be exploited to process some information in order to reduce the communication overhead and the computational burden. It also reduces latency as certain operations can already be performed in parallel at the nodes. Here, we first give a brief introduction of general ADMM formulation followed by our proposed reformulations of the problem in (\ref{eq:OptMod}) according to ADMM framework.\par
Consider the following constrained optimization problem with linear constraints over two separable functions in two variables $\mathbf{u}$ and $\mathbf{z}$
\begin{equation}
\begin{gathered}
\label{eq:ADMM}
{\text{arg}}\mathop {\min }\limits_{\mathbf{u},\mathbf{z}} f(\mathbf{u}) + g(\mathbf{z}) \hfill \\
 s.t.\,\,\,\,\,\,\,\,\mathbf{G} \mathbf{u} + \mathbf{H} \mathbf{z} = \mathbf{c},
\end{gathered}
\end{equation}
where $\mathbf{G}$, $\mathbf{H}$, and $\mathbf{c}$ are the matrices and vector of appropriate dimensions that establish the constraints on the variables $\mathbf{u}$ and $\mathbf{z}$.
The augmented Lagrangian function of the above problem becomes
\begin{equation}
\label{eq:AugLag}
 \mathcal{L}\left(\mathbf{u},\mathbf{z},{\boldsymbol{\sigma}} \right) =  \left\{
 \begin{gathered}
 \begin{aligned} & f(\mathbf{u}) + g(\mathbf{z}) + \left\langle {\boldsymbol{\sigma }},{\mathbf{G}{\mathbf{u}} + \mathbf{H} \mathbf{z} - \mathbf{c}} \right\rangle \\
  + & \frac{\beta }{2}\left\|{\mathbf{G}{\mathbf{u}} + \mathbf{H} \mathbf{z} - \mathbf{c}}\right\|_2^2\hfill \end{aligned} \end{gathered} \right\}
\end{equation}
where ${\boldsymbol{\sigma}}$ is the dual variable, $\beta$ is the augmented Lagrangian parameter, and $\left< \cdot, \cdot \right>$ denotes the inner product of vectors.\par
The ADMM solution to the above problem is obtained by iteratively minimizing the augmented Lagrangian function with respect to both the variables $\mathbf{u}$ and $\mathbf{z}$ in an alternating fashion in addition to updating the dual variable each iteration. Accordingly, after the $k^\text{th}$ iteration, the ADMM variable updates consist of \cite{boyd_distributed_2011}
\begin{equation}
\label{eq:VarUpdt}
 \begin{aligned}
 \mathbf{u}^{\left(k+1 \right)}:= & {\text{arg}}\mathop {\min }\limits_{\mathbf{u}} \mathcal{L} \left(\mathbf{u},\mathbf{z}^{\left( k \right)},{\boldsymbol{\sigma }}^{\left( k \right)} \right) \\
 \mathbf{z}^{\left(k+1 \right)}:= & {\text{arg}}\mathop {\min }\limits_{\mathbf{z}} \mathcal{L} \left(\mathbf{u}^{\left( k \right)},\mathbf{z},{\boldsymbol{\sigma }}^{\left(k \right)} \right)\\
  {\boldsymbol{\sigma}}^{\left( k+1 \right)}:= & {\boldsymbol{\sigma }}^{\left ( k \right)} + {\beta} \left(\mathbf{G} \mathbf{x}^{\left( k+1 \right)} + \mathbf{H} \mathbf{z}^{\left( k+1 \right)} - \mathbf{c} \right)
 \end{aligned}   
\end{equation}
Embracing ADMM framework, we propose two different formulations for (\ref{eq:OptMod}). By introducing a new variable $\mathbf{x_{G}} \in \mathbb{R}^{N \times 1}$ representing the magnitude of the global image, both the formulations will have the same objective function of minimizing the sum of the least square terms with respect to local images, in addition to minimizing the $l_1$ norm of the global image. The formulations differ in the constraints which define the relation between the global and local images. We consider the magnitude of the images as our optimization variables assuming that the phases are estimated in a previous step. Specifically, we assume that we have estimated $\mathbf{\Theta}_{q} \in \mathbb{C}^{N \times N}$, the diagonal matrix containing the phase of all pixels of local image over its diagonal such that $\tilde{\mathbf{x}}_{q}=\mathbf{\Theta}_{q} {\mathbf{x}}_{q}$. For ease of notation, from now on we will consider the matrix $\mathbf{\Theta}_{q}$ included in the measurement matrix $\mathbf{A}_{q}$. The details regarding the estimation of $\mathbf{\Theta}_{q}$ will be discussed in the next section. Accordingly, with reference to (\ref{eq:ADMM}), our first variable is $\mathbf{x} \in \mathbb{R}^{QN \times 1}$ containing the magnitude of all local images $\mathbf{x} = \{\mathbf{x}_{q}\}_{q=1}^{Q}$, and the second variable represents the magnitude of the global image $\mathbf{x_{G}}$. Consequently, our objective function will be $f(\mathbf{x}) = \sum_{q=1}^{Q}{ \left\| \mathbf{y}_{q}-\mathbf{A}_{q}\mathbf{x}_{q} \right\|_{2}^{2}}$ and $g(\mathbf{x_{G}}) = \left\|{\mathbf{{x}_{G}}}\right\|_{1}$.\par
In the sequel, we will provide our proposed aforementioned formulations and their solutions in terms of variable updates according to (\ref{eq:VarUpdt}).
\subsection{Consensus ADMM (CADMM)}
\label{sec:CADMM}
As the name suggests, by posing the problem according to this formulation, we pursue a solution which, at optimum, provides a global image on which all clusters reach a consensus. Consequently, the constraints, in this case, are defined to impose this relationship between the global and local images. Additionally, as mentioned earlier and by following our paper \cite{hu_widely-distributed_2021}, we impose the $l_1$-norm function to promote sparse global image solution. The problem becomes 
%
\begin{equation}
\label{eq:CADMM_problem}
\begin{gathered}
\underset{\mathbf{x},\mathbf{{x}_{G}}}{\text{arg}\mathop{\min}}\,\, \, \, \,\sum_{q=1}^{Q}{ \frac{\mathrm{\mu}}{2}\left\| \mathbf{y}_{q}-\mathbf{A}_{q}\mathbf{x}_{q} \right\| _{2}^{2}} + \lambda \left\|{\mathbf{{x}_{G}}}\right\|_{1}\,\,\,\,\,
\hfill \\
\,\,\,\,\, \, \, \, \, \,\, \, \, \,\, \, \, \, \, \, \,\, \, \, \,  s.t.\, \, \,\, \, \, \,\, \, \,\, \, \, \,\, \, \,\, \, \, \,\mathbf{x}_{q}-\mathbf{{x}_{G}}=\mathbf{0}\,\,\,\,\,\,\,\forall q.\,\,\,\,\,\,\,\hfill \\
\end{gathered}
\end{equation}
where $\lambda$ and $\mu$ are positive hyperparameters set to penalize less sparse global image solutions and trade-off the data fidelity term, respectively.
%
Note that the $Q$ constraints in (\ref{eq:CADMM_problem}) can be written in the form of the constraint in (\ref{eq:ADMM}) by having $\mathbf{G} = \mathbf{I}_{QN}$, $\mathbf{u}=\mathbf{x}$, $\mathbf{H}=-[\mathbf{I}_{N}, \mathbf{I}_{N}, \cdots \mathbf{I}_{N}]^{T}$ of the size $QN \times N$, $\mathbf{z}=\mathbf{x_{G}}$, and $\mathbf{c} = \mathbf{0}$ of size $QN \times 1$.\par
%
As indicated in (\ref{eq:VarUpdt}), the solution of (\ref{eq:CADMM_problem}) can be obtained by alternately minimizing its associated augmented Lagrangian with respect to $\mathbf{x}$, $\mathbf{{x}_{G}}$, and the dual variable $\boldsymbol{\sigma}$. The augmented Lagrangian is
\begin{equation}
\label{eq:aug_lag_C}
\begin{gathered}
  \mathcal{L}\left( {{\mathbf{x}},{\mathbf{{x}_{G}}},{\boldsymbol{\sigma }}} \right) =  \\
  \left\{
  \begin{gathered}
  \sum\limits_{q = 1}^{Q}
  \begin{gathered} 
  \left\{
  \frac{\mu }{2}\left\|{{{\mathbf{y}}_{q}} - {{\mathbf{A}}_{q}}{{\mathbf{x}}_{q}}} \right\|_2^2 + \left\langle {{{\boldsymbol{\sigma }}_{q}},{{\mathbf{x}}_{q}}-{\mathbf{{x}_{G}}} } \right\rangle  
  + \frac{\beta }{2}\left\| { {{\mathbf{x}}_{q}}-{\mathbf{{x}_{G}}} } \right\|_2^2 \right\} \end{gathered} 
   \\ +\lambda {\left\| {\mathbf{{x}_{G}}} \right\|_1}\end{gathered} \right\},
\end{gathered} 
\end{equation}
where ${\boldsymbol{\sigma }}_{q} \in \mathbb{R}^{N \times 1}$ is the block of values of the dual variable ${\boldsymbol{\sigma }} \in \mathbb{R}^{QN \times 1}$ which corresponds to the local image $\mathbf{x}_{q}$.\par
The resulting updates of each variable according to CADMM formulation are provided hereinafter in details.
\\
\subsubsection{Update of $\mathbf{x}$ (Local Images)}
Let $\mathbf{{x}_{G}}^{\left(k\right)}$ and $\boldsymbol{\sigma }^{\left(k\right)}$ denote the values of $\mathbf{{x}_{G}}$ and $\boldsymbol{\sigma }$ after the $k^{\text{th}}$ iteration. Since $\mathcal{L}\left( \mathbf{x},\mathbf{x_{G}},\boldsymbol{\sigma} \right)$ in (\ref{eq:aug_lag_C}) is decomposable with respect to ${\mathbf{x}_{q}}$, the updated ${\mathbf{{x}}}^{\left(k+1\right)}$ can be obtained by updating all local images $\mathbf{{x}}_{q}^{\left(k+1\right)}$ for $q=1,\ldots,Q$ in parallel as
\begin{equation}
\begin{gathered}
\begin{aligned}
\label{eq:updt_x_CADMM}
  {\mathbf{x}}_{q}^{\left(k + 1\right)} &= {\text{arg}}\mathop {\min }\limits_{{{\mathbf{x}}_{q}}} {\mathcal{L}}\left( {{{\mathbf{x}}_{q}};{{\mathbf{{x}_{G}}}^{\left(k \right)}},{\boldsymbol{\sigma }}_{q}^{\left(k\right)}} \right) \\
    &= 
    {\text{arg}}\mathop {\min }\limits_{{{\mathbf{x}}_{q}}}
    \left\{ \begin{gathered}
    \frac{\mu }{2}\left\| {{{\mathbf{y}}_{q}} - {{\mathbf{A}}_{q}}{{\mathbf{x}}_{q}}} \right\|_2^2 + {\boldsymbol{\sigma}_{q}^{\left(k \right)}}^{T}  {{\mathbf{x}}_{q}} \\  + \frac{\beta}{2}\left\| {{{\mathbf{x}}_{q}} - {{\mathbf{{x}_{G}}}^{\left(k \right)}}} \right\|_2^2 
    \end{gathered} \right\},
\end{aligned}
\end{gathered}
\end{equation}
The problem in (\ref{eq:updt_x_CADMM}) is differentiable with respect to $\mathbf{x}_{q}$ and the $(k+1)^{\text{st}}$ update can be obtained in a closed-form by letting $\nabla_{\mathbf{x}_{q}}\mathcal{L} =\mathbf{0}$ resulting in
\begin{equation}
\label{eq:updt_x_CADMM_CF}
    {\mathbf{x}}_{q}^{\left(k + 1\right)} = \left( \mu \mathbf{A}_{q}^{H}\mathbf{A}_{q}+\beta \mathbf{I}_{N} \right)^{-1} \left( \mu \mathbf{A}_{q}^{H}\mathbf{y}_{q}+\beta {{\mathbf{{x}_{G}}}^{\left(k \right)}} - {\boldsymbol{\sigma }}_{q}^{\left(k\right)} \right),
\end{equation}
Note that the inverse in (\ref{eq:updt_x_CADMM_CF}) is possible since $\left(\mu \mathbf{A}_{q}^{H}\mathbf{A}_{q}+\beta \mathbf{I}_{N}\right)$ is a positive definite matrix.
\\
\subsubsection{Update of $\mathbf{x_{G}}$ (Global Image)}
For the global image update, following the ADMM framework, we consider
\begin{equation}
\begin{gathered}
\begin{aligned}
\label{eq:updt_xG_CADMM}
{\mathbf{{x}}^{\left(k + 1\right)}_\mathbf{G}} &={\text{arg}}\mathop {\min }\limits_{{\mathbf{x_{G}}}}\mathcal{L}\left(\mathbf{x_{G}};\mathbf{{x}}^{\left(k+1 \right)},\boldsymbol{\sigma }^{\left(k \right)} \right) \hfill \\
 &= {\text{arg}}\mathop {\min }\limits_{{\mathbf{{x}_{G}}}} \left\{ \begin{aligned}\lambda \left\| {\mathbf{x_{G}}} \right\|_1 +  \sum_{q=1}^{Q} {\boldsymbol{\sigma}_{q}^{\left(k \right)}}^{T} {\mathbf{{x}_{G}}} \hfill\\
 + \frac{\beta}{2} \sum_{q=1}^{Q}\left\| {\mathbf{{x}}_{q}^{\left(k+1 \right)} - {{\mathbf{{x}_{G}}}}} \right\|_2^2
 \end{aligned}  \right\}.  
\end{aligned}
\end{gathered}
\end{equation}
This objective function above involves information from all $Q$ clusters and is not decomposable with respect to $\mathbf{{x}_{G}}$. It further involves a non-differentiable function $\left\| \mathbf{{x}_{G}} \right\|_1$. Thus, it can neither be parallelized nor solved in a closed form like (\ref{eq:updt_x_CADMM}). As a result, it is more suitable for the global image update to be carried out in a central processor after collecting local updates calculated at the distributed clusters. Moreover, for the subsequent update of local images, global image needs to be broadcast to all the clusters. Alternatively, if the global image update were to be carried out at distributed clusters, a fully meshed communication network would be needed to exchange all local updates among the $Q$ clusters. Later in \ref{sec:SolTech}, we will show how to solve (\ref{eq:updt_xG_CADMM}) in the central node. 
\\
\subsubsection{Update of ${\boldsymbol{\sigma}}$ (Dual Variable)}
After updating the global image, the dual variable can be updated by
\begin{equation}
\label{eq:updt_sig_CADMM}
\boldsymbol{\sigma }^{\left(k + 1\right)} = \boldsymbol{\sigma}^{\left(k \right)} + \beta \left( {\mathbf{{x}}}^{\left(k + 1\right)} - \mathbf{1}_{Q} \otimes {{\mathbf{x}}}_{\mathbf{{G}}}^{\left(k + 1\right)} \right), 
\end{equation}
The Kronecker product is used to replicate the global image to the same size of the vectors ${\mathbf{{x}}}$ and $\boldsymbol{\sigma}$. Since (\ref{eq:updt_sig_CADMM}) is decomposable, it can be carried out in parallel as well as local images updates. Instead, it is more convenient for the dual variable to be updated in the central node subsequent to that of global images and broadcast to the distributed clusters for the next update of the local image.
\subsection{Sharing ADMM (SADMM)}
\label{sec:SADMM}
Under this formulation, we impose a different constraint in the optimization problem to explore a different relationship between local images and global image. The constraint is set such as the reconstructed global image is the average of all local images. Accordingly, the problem becomes
\begin{equation}
\label{eq:SADMM}
\begin{gathered}
\underset{\mathbf{x},\mathbf{{x}_{G}}}{\text{arg}\mathop{\min}}\,\, \, \, \,\sum_{q=1}^{Q}{ \frac{\mathrm{\mu}}{2}\left\| \mathbf{y}_{q}-\mathbf{A}_{q}\mathbf{x}_{q} \right\| _{2}^{2}} + \lambda \left\|{\mathbf{{x}_{G}}}\right\|_{1}\,\,\,\,\,
\hfill \\
\,\,\,\,\, \, \, \, \, \,\, \, \, \,\, \, \, \, \, \, \,\, \, \, \,  s.t.\, \, \,\, \, \, \,\, \, \,\, \, \, \,\, \, \,\, \, \, \,\bar{\mathbf{x}}-{\mathbf{x}_\mathbf{{G}}}=\mathbf{0}\,\,\,\,\,\,\,\,\,\,\,\,\,\,\hfill \\
\end{gathered}
\end{equation}
where $\bar{\mathbf{x}} = \sum_{q=1}^{Q} \mathbf{x}_{q}$ contains the sum of magnitudes of local images. Note that the size of constraints is reduced to the size of a single image instead of $Q$ images in the consensus formulation. The nomenclature stems from the constraint above since the global image is considered a shared combination of all local images. We can again write the constraint of (\ref{eq:SADMM}) in the form of the constraint in (\ref{eq:ADMM}) by having $\mathbf{G} =  [\mathbf{I}_{N}, \mathbf{I}_{N}, \cdots \mathbf{I}_{N}]$ of size $N \times QN$,  $\mathbf{u} = \mathbf{x}$, $\mathbf{H}=-\mathbf{I}_{N} $, $\mathbf{z} = \mathbf{x_{G}}$, and $\mathbf{c} = \mathbf{0}$.\par
The augmented Lagrangian of (\ref{eq:SADMM}) can then be written as
\begin{equation}
\label{eq:aug_lag_S}
\begin{gathered}
  \mathcal{L}\left({{\mathbf{x}},{\mathbf{{x}_{G}}},{\boldsymbol{\sigma }}} \right) = \left\{ \begin{gathered} \sum\limits_{q = 1}^{Q} \frac{\mu }{2}\left\| {{{\mathbf{y}}_{q}} - {{\mathbf{A}}_{q}}{{\mathbf{x}}_{q}}} \right\|_2^2 + {\lambda}{\left\| {\mathbf{{x}_{G}}} \right\|_1} \\ \hfill+\left\langle {{{\boldsymbol{\sigma }}},{\mathbf{\bar{x}}} - {\mathbf{{x}_{G}}}} \right\rangle + \frac{\beta }{2}\left\| {{\mathbf{\bar{x}}} - {\mathbf{{x}_{G}}}} \right\|_2^2  \end{gathered} \right\}.
\end{gathered} 
\end{equation}
Note that since the number of constraints are reduced, the dual variable ${\boldsymbol{\sigma}}$ has a size ${N \times 1}$ instead of ${QN \times 1}$ as in CADMM. Next, we provide the variable updates due to SADMM formulation. 
\\
\subsubsection{Update of $\mathbf{x}$ (Local Images)}
Unlike the consensus case, the augmented Lagrangian function (\ref{eq:aug_lag_S}) is not directly decomposable into $Q$ terms because of the sum variable $\bar{\mathbf{x}}$ inside the augmented quadratic term. However, we show here that it is still possible to solve for each local image variable $\mathbf{x}_{q}$ in parallel. Similar to (\ref{eq:updt_x_CADMM}), we use the values of ${\mathbf{{x}}^{\left(k \right)}_\mathbf{{G}}}$ and ${{\boldsymbol{\sigma }}_{q}^{\left(k\right)}}$ in order to solve for $\mathbf{x}_{q}$ at the $(k+1)^\text{st}$ iteration. However, since we have also $\bar{\mathbf{x}}$ in (\ref{eq:aug_lag_S}), we fix also all other local image variables to $\mathbf{x}_{i}^{\left(k \right)} \ \forall i \neq q$. Let $ \bar{\mathbf{x}}_{q}^{\left(k \right)} = \sum_{\left(i \neq q\right)} {\mathbf{x}}_{i}^{\left(k \right)} = \bar{\mathbf{x}}^{\left(k \right)} - {\mathbf{x}}_{q}^{\left(k \right)}$. Consequently, the $q^{\text{th}}$ local image update can be obtained by
\begin{equation}
\begin{aligned}
\label{eq:updt_x_SADMM}
  {\mathbf{x}}_{q}^{\left(k + 1\right)} &= {\text{arg}}\mathop {\min }\limits_{{{\mathbf{x}}_{q}}} {\mathcal{L}}\left( {{{\mathbf{x}}_{q}};\bar{\mathbf{x}}_{q}^{\left(k \right)},{\mathbf{{x}}^{\left(k \right)}_\mathbf{{G}}},{\boldsymbol{\sigma }}^{\left(k\right)}} \right) \hfill \\
    &= 
    {\text{arg}}\mathop {\min }\limits_{{{\mathbf{x}}_{q}}} \left\{ \begin{gathered}\frac{\mu }{2}\left\| {{{\mathbf{y}}_{q}} - {{\mathbf{A}}_{q}}{{\mathbf{x}}_{q}}} \right\|_2^2  + {\boldsymbol{\sigma}^{\left(k \right)}}^{T}  {{\mathbf{x}}_{q}} \\ + \frac{\beta}{2}\left\| {{{\mathbf{x}}_{q}} + \bar{\mathbf{x}}_{q}^{\left(k \right)} - {\mathbf{{x}}^{\left(k \right)}_\mathbf{{G}}}}  \right\|_2^2
    \end{gathered} \right\}\hfill \\ 
    &= 
    {\text{arg}}\mathop {\min }\limits_{{{\mathbf{x}}_{q}}} \left\{ \begin{gathered}\frac{\mu }{2}\left\| {{{\mathbf{y}}_{q}} - {{\mathbf{A}}_{q}}{{\mathbf{x}}_{q}}} \right\|_2^2  + {\boldsymbol{\sigma}^{\left(k \right)}}^{T}  {{\mathbf{x}}_{q}} \\ + \frac{\beta}{2}\left\| {{{\mathbf{x}}_{q}} + \bar{\mathbf{x}}^{\left(k \right)} - {\mathbf{x}}_{q}^{\left(k \right)} - {\mathbf{{x}}^{\left(k \right)}_\mathbf{{G}}}}  \right\|_2^2
    \end{gathered}  \right\}\hfill
\end{aligned}
\end{equation}
Now similar to (\ref{eq:updt_x_CADMM}), the problem in (\ref{eq:updt_x_SADMM}) is fully differentiable with respect to $\mathbf{x}_{q}$ and the $k+1^\text{st}$ update can be obtained in  the closed-form 
\begin{equation}
\label{eq:updt_x_SADMM_CF}
{\mathbf{x}}_{q}^{\left(k + 1\right)} = \resizebox{.85\hsize}{!}{$ \left( \mu \mathbf{A}_{q}^{H}\mathbf{A}_{q}+\beta \mathbf{I}_{N} \right)^{-1} \left( \mu \mathbf{A}_{q}^{H}\mathbf{y}_{q}+\beta \left({\mathbf{{x}}^{\left(k \right)}_\mathbf{{G}}}-{\bar{\mathbf{x}}_{q}^{\left(k\right)}} \right) - {\boldsymbol{\sigma }}^{\left(k\right)} \right)$}.
\end{equation}
From (\ref{eq:updt_x_SADMM_CF}), we can observe that in SADMM, the $q^{\text{th}}$ local image update requires the previous state ${\mathbf{x}}_{q}^{\left(k \right)}$ and the sum of all other local images earlier updates $\bar{\mathbf{x}}^{\left(k \right)}$ in addition to the global image update ${{\mathbf{{x}}_{G}}^{\left(k \right)}}$, and the dual variable update ${\boldsymbol{\sigma}_{q}^{\left(k \right)}}$. This suggests the need for extra memory with respect to CADMM to track the previous state at each cluster. Additionally, the distributed clusters will need to receive each other updates. This can be broadcast by the central node subsequent to the update of global image and dual variables. The central node will have such values any way since they are needed for the global image update. Although the exchanged information between the central node and the distributed clusters in SADMM seems to be more than CADMM, the size of those variables to be exchanged is still less than CADMM by a factor of $(Q+2)/3$ due to the reduced size of the dual variable in SADMM. This provides a significant reduction in communication bandwidth requirements between the central node and the sensors especially for a large number of distributed sensors.
\\
\subsubsection{Update of $\mathbf{{x}_{G}}$ (Global Image)}
Similar to the global image update in Section \ref{sec:CADMM}, after collecting the local images updates from the distributed sensors, the global image update for the sharing formulation is obtained by minimizing the augmented Lagrangian with respect to ${\mathbf{{x}_{G}}}$ as follows
\begin{equation}
\begin{gathered}
\label{eq:updt_xG_SADMM}
{\mathbf{{x}}^{\left(k+1 \right)}_\mathbf{{G}}} =\resizebox{.85\hsize}{!}{$ \mathrm{arg}\mathop{\min} \limits_{{\mathbf{x_{G}}}}\left\{ \lambda \left\| {\mathbf{x_{G}}} \right\|_1 + \frac{\beta}{2}\left\| {\mathbf{x_{G}}} - \bar{\mathbf{x}}^{\left(k + 1\right)} \right\|_{2}^{2} +{\boldsymbol{\sigma}^{\left(k \right)}}^{T} {\mathbf{{x}_{G}}}\right\}.$}
\end{gathered}
\end{equation}
Again, we here assume that both the global image and the dual variable is calculated at the central node. As a result, the sum of the local images $\bar{\mathbf{x}}^{\left(k \right)}$ is calculated directly at the central node following the local updates collection needed for the global image update. The solution of (\ref{eq:updt_xG_SADMM}) will be detailed later in the section.
\\
\subsubsection{Update of ${\boldsymbol{\sigma }}$ (Dual variable)}
The dual variable update then is a straight forward step of the ADMM algorithm
\begin{equation}
\label{eq:updt_sig_SADMM}
\boldsymbol{\sigma }^{\left(k + 1\right)} = \boldsymbol{\sigma}^{\left(k \right)} + \beta \left( \bar{\mathbf{x}}^{\left(k + 1\right)} - {{\mathbf{x}}}_{\mathbf{{G}}}^{\left(k + 1\right)} \right) 
\end{equation}
To summarize, despite the performance of each formulation in terms of image quality which will be examined in the next section, SADMM provides an alternative processing architecture with respect to CADMM in which: 1) extra memory is needed at the distributed clusters for local image updates, 2) communication overhead between the distributed clusters and the central node is reduced by a factor of $(Q+2)/3$ due to the size difference of the constraints. In both CADMM and SADMM, local images can be updated in parallel at each cluster node and communicated back to the central node. The central node in turns updates both the global image and the dual variable. Subsequently, both the global image and dual variable updates are broadcast back to the distributed clusters in the case of CADMM in addition to the sum of the previous local images in case of SADMM in order to calculate the next local image updates. Other comparisons and performance metrics such as image reconstruction quality and convergence rate will be provided later in Section \ref{sec:Perf_Eval}.
\subsection{Solution Techniques}
\label{sec:SolTech}
Considering the above formulations, in this section, we provide the techniques used to solve the sub-problems for local and global images updates. Additionally, we provide the stopping criteria adopted to terminate both algorithms. Lastly, we show how to obtain the phases of complex-valued local images prior to ADMM iterations.
\\
\subsubsection{Variable Updates}
The update of local images employs a matrix inversion in a closed-form solution. However, due to the large size of the problem, it needs to be solved iteratively using a numerical procedure. In particular, we carry out the inversion in the local update using conjugate gradient (CG) method \cite{shewchuk1994introduction}. Being a numerical method, the output of CG is a complex-valued imaged since both the measurements $\mathbf{y}_{q}$ and the forward model $\mathbf{A}_{q}$ are complex. However, the optimization is carried out over the real-valued magnitude of the images where the phase of the images is included in the measurement matrix. Therefore, subsequent to the update of the local image using CG, a projection of the resulting complex image on the real positive orthant is applied to obtain the magnitude of the local image. This projection implicitly states that the phase of the complex-valued output of CG is regarded as a numerical phase error.\par
The global image on the other hand, requires solving a LASSO like optimization problem which can be solved using a proximal gradient method. In our numerical experiments, we used the accelerated proximal gradient \cite{beck2009fast} to calculate the global image updates.
\\
\subsubsection{Stopping Criteria}
Variable updates are repeated until termination which is decided upon comparing the values of the primal and dual residuals with their corresponding feasibility tolerances $\epsilon_{pri}$ and $\epsilon_{dual}$ respectively. Following the definitions of the residuals and the stopping criteria brought up in \cite{boyd_distributed_2011}, let $\boldsymbol{\eta}_{pri}$ and $\boldsymbol{\eta}_{dual}$ denote the primal and dual residuals respectively. The dual residual is defined over the successive updates of global image variable. Hence, it is the same for both CADMM and SADMM. Accordingly, at the $k^{\text{th}}$ iteration, the dual residual for both formulations is given by
\begin{equation}
\label{eq:resid_dual}
\begin{aligned}
\boldsymbol{\eta }^{\left(k\right)}_{dual} = \beta \left( {{\mathbf{x}}}_{\mathbf{{G}}}^{\left(k + 1\right)} - {{\mathbf{x}}}_{\mathbf{{G}}}^{\left(k \right)}\right).
\end{aligned}
\end{equation}
On the other hand, the primal residual measures the constraint satisfaction and takes a different form in the two formulations. For CADMM, the primal residual is given by
\begin{equation}
\label{eq:resid_pri_C}
\begin{aligned}
&\boldsymbol{\eta }^{\left(k\right)}_{pri}= {\mathbf{{x}}}^{\left(k + 1\right)} - \mathbf{1}_{Q} \otimes {{\mathbf{x}}}_{\mathbf{{G}}}^{\left(k + 1\right)}.
\end{aligned}
\end{equation}
Similarly, in SADMM the primal residual is
\begin{equation}
\label{eq:resid_pri_S}
\begin{aligned}
&\boldsymbol{\eta }^{\left(k\right)}_{pri}= \bar{\mathbf{{x}}}^{\left(k + 1\right)} - {{\mathbf{x}}}_{\mathbf{{G}}}^{\left(k + 1\right)}.
\end{aligned}
\end{equation}
The feasibility tolerances can be chosen based on an absolute tolerance $\epsilon_{abs}$ and a relative tolerance $\epsilon_{rel}$. Similar to the primal and dual residuals, the feasibility tolerances indicate non-identical definitions depending on the formulation due to the different constraints. In CADMM, they are given by
\begin{equation}
\begin{aligned}
\label{eq:tol_C}
&\epsilon_{pri}= \sqrt{QN} \ \epsilon_{abs} + \epsilon_{rel} \ \text{max} \left\{ \left\| {\mathbf{{x}}}^{\left(k \right)} \right\|_2,  \sqrt{Q}\left\| {{\mathbf{x}}}_{\mathbf{{G}}}^{\left(k \right)} \right\|_2 \right \}\\
&\epsilon_{dual} = \sqrt{QN} \ \epsilon_{abs} + \epsilon_{rel} \left\| {\boldsymbol{\sigma }}^{\left(k \right)} \right\|_2.
\end{aligned}
\end{equation}
Likewise, for SADMM, feasibility tolerances are 
\begin{equation}
\begin{aligned}
\label{eq:tol_S}
&\epsilon_{pri}= \sqrt{N} \ \epsilon_{abs} + \epsilon_{rel} \ \text{max} \left\{ \left\| \bar{\mathbf{{x}}}^{\left(k \right)} \right\|_2, \left\| {{\mathbf{x}}}_{\mathbf{{G}}}^{\left(k \right)} \right\|_2 \right \}\\
&\epsilon_{dual} = \sqrt{N} \ \epsilon_{abs} + \epsilon_{rel} \left\| {\boldsymbol{\sigma }}^{\left(k \right)} \right\|_2.
\end{aligned}
\end{equation}
Lastly, for both CADMM and SADMM, the algorithm is terminated either when a defined maximum number of iterations is reached or when both the ensuing inequalities are satisfied
\begin{equation}
\begin{aligned}
\label{eq:stop}
& \left\| \boldsymbol{\eta }^{\left(k\right)}_{pri} \right\|_2 \leqslant \epsilon_{pri}\\
& \left\| \boldsymbol{\eta }^{\left(k\right)}_{dual} \right\|_2 \leqslant \epsilon_{dual},
\end{aligned}
\end{equation}
where the variables in the above inequalities are calculated according to the definitions of the corresponding formulation.
\\
\subsubsection{Phase Matrix $\mathbf{\Theta}$}
As mentioned earlier, we assume that the phase of local images is already provided prior to carrying out the optimization algorithms. Our proposed imaging methods can be considered partially non-coherent imaging method since the phases are only used within the data-fidelity term in the objective function. Thus, a coarse estimated phase of local images is sufficient for our algorithms to perform satisfactorily. Therefore, we use the phase of the images obtained by back-projection for each cluster as an estimate of the phase of local images. Accordingly, for each cluster $q$, the diagonal matrix containing the phase of all pixels of its local image is constructed as
\begin{equation}
    \mathbf{\Theta}_{q} = \text{diag}\left\{\exp{(j[\angle{(\mathbf{A}_{q}^{H} \mathbf{y}_{q})])}}\right\}.
\end{equation}
%
\section{Performance Evaluation}
\label{sec:Perf_Eval}
In this section, we validate and evaluate the performance of the algorithms proposed in Section \ref{sec:ADMM} to reconstruct radar images using distributed sensor clusters. To achieve this goal, we use the publicly available civilian vehicle dome (CV domes) data-set which offers simulated scattering data of civilian vehicle facet models. Although the data-set is originally intended to simulate circular synthetic aperture radars, a particular configuration of WSAR, it can also be used to simulate a mono-static distributed radar sensors system.\par
First, we give a brief introduction of the data-set and its parameters. Consequently, we define the performance metrics used in our evaluation to compare both algorithms. Finally, we evaluate our algorithms on three different scenarios of practical relevance for several applications. The scenarios are realized by different combinations of full/limited views and full/limited bandwidth measurements as we will show later in this section.
\subsection{Data-set Introduction}
CV Domes data-set contains simulated high-frequency scattering data of ten civilian vehicles. For each model, an X-band electromagnetic mono-static scattering is simulated in a far-field scenario. Scattered waves are simulated with full polarization over an azimuth extent of $360 \degree$ where $16$ viewing angles per degree of azimuth are considered. Similarly, data are simulated over the range of elevation angles from $30\degree$ to $60 \degree$. For each azimuth and elevation viewing angle tuple, $512$ frequency samples of complex-valued scattering coefficients centered at $9.6$ GHz and spanning a bandwidth of approximately $5.35$ GHz are provided. The range information of those frequency measurements is compressed already resulting in what is usually referred to as phase history.\par
\subsection{Performance Metrics}
As discussed in the previous section, the difference in problem formulation between CADMM and SADMM has induced slightly different system implementation features in terms of memory requirements and communication bandwidth. Additionally, to compare the performance 
of the proposed algorithms, the following aspects are considered.
\begin{enumerate}
    \item Convergence rate: it can be assessed by the number of iterations needed to reach the stopping criteria that is defined identically for both algorithms.
    \item Computational complexity: the main computational burden of both algorithms lies in the local image updates (\ref{eq:updt_x_CADMM_CF}) and (\ref{eq:updt_x_SADMM_CF}), where the matrix inversion term is present. Due to their large size, the measurement matrices $\mathbf{A}_{q}$ are realized through matrix operators based on two-dimensional non-uniform Fast Fourier transform (2D NuFFT) \cite{fessler_nonuniform_2003}. Moreover, the inversion step is carried out numerically using CG as mentioned earlier. Thus, while the complexity of both algorithms seems to be equivalent, the convergence of CG highly depends on the other variables in the update formulas (\ref{eq:updt_x_CADMM_CF}) and (\ref{eq:updt_x_SADMM_CF}). As a result, the comparison solely in terms of the number of iterations is not indicative since a single iteration in each of the algorithms may realize a different cost. Accordingly, computational complexity can be measured by calculating the total processing time spent until termination.
    \item Image reconstruction quality: the data-set does not contain a reference image with which a comparison can be made in order to evaluate the quality of reconstructed images. Correspondingly, we use the image entropy as a quantitative metric to assess the image quality being a measure of its sharpness or constituent randomness. The smaller the image entropy the sharper is the reconstructed image and vice versa. The entropy of an image is calculated in bits after intensity saturation for the values beyond a desired dynamic range in the dB scale followed by image translation to the gray-scale. Consequently, a randomly generated image would have an entropy equal or close to $8$ bits. Additionally, as a subjective measure of images quality, the images reconstructed utilizing full aperture and full bandwidth measurement could act as a visual reference for the other scenarios when the aperture and/or the bandwidth measurements are reduced.
\end{enumerate}
\subsection{Experiments}
\label{sec:experiments}
\begin{figure*}[t!]
	\centering
	\includegraphics[width=7 in]{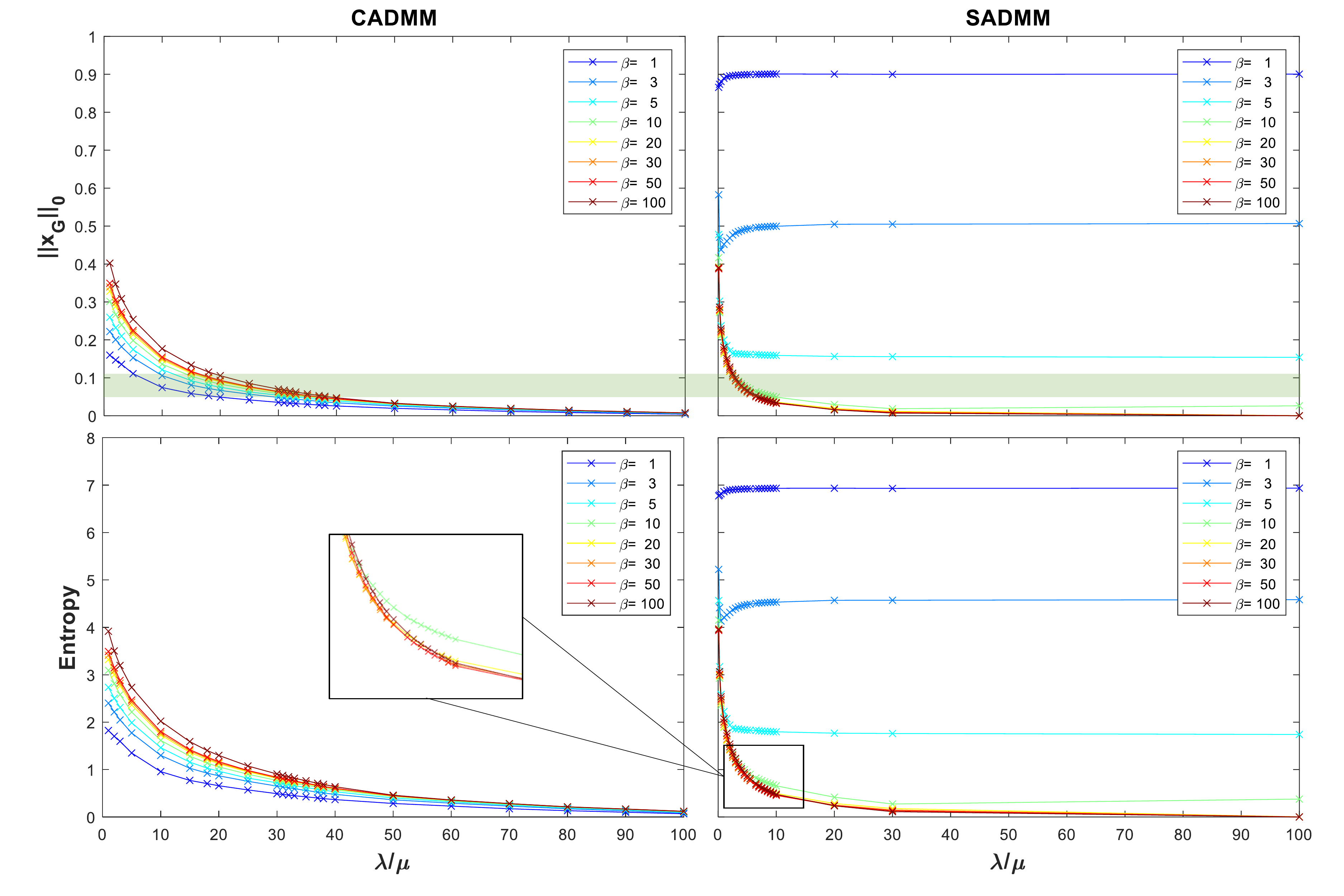}
	\caption{Hyperparameters sweep for 'Jeep99' data-set (FVFB); normalized sparsity (top row), image entropy (bottom row)}
	\label{fig:FVFB_ParSw}
\end{figure*}
In this subsection, we illustrate the reconstructed images by CADMM and SADMM using the simulated data of two different vehicle models differing in type and geometry. The first is of a Jeep Cherokee (SUV) 'Jeep99', while the second is of a Toyota Tacoma (Pick-up) 'Tacoma'. We consider image reconstruction utilizing the data-set according to the following scenarios:
\begin{enumerate}
    \item Full aperture views measurements: for a general validation of both algorithms, the full $360 \degree$ aperture measurements of the entire available bandwidth is considered.
    \item Full views and limited bandwidth measurements: limited frequency samples of the full aperture measurements are considered for image reconstruction realizing a typical use case of WSAR imaging.
    \item Limited views and limited bandwidth measurements: assuming a distributed system of radar sensors illuminating the scene according to a time division multiplexing (TDM) scheme, limited frequency samples of limited aperture measurements are considered for images reconstruction.
\end{enumerate}
%
For all experiments, measurements are taken at a fixed elevation angle of $30 \degree$ and with 'HH' polarization. Moreover, all measurements are impaired with a white Gaussian noise realizing a signal to noise ratio (SNR) of $15$ dB. A fine grid of $256$ cells in both range and cross-range directions ($7$ meter-long each) is used resulting in a total number of $N =65536$ pixels. Additionally, images are reconstructed considering an elevated image plane at $1$ meter from the ground level. This renders the projection of layover-ed elements to be mostly contained within vehicles outlines and permits a better visual interpretation.\par
The choice of $\mu$, $\beta$, and $\lambda$ for both CADMM and SADMM is made through a parameter sweep guided by normalized image sparsity and image entropy as performance metrics. The normalized sparsity considered is the percentage of the non-zero pixels in the image. Hyperparameters used to reconstruct the illustrated images throughout this section are those that guarantee a similar degree of sparsity for both CADMM and SADMM images at a lower entropy value. Also, we tried to pick the parameters where $\beta$ is as close as possible in both methods for a fair convergence comparison. Given the considered scene size and typical dimensions of a vehicle, a sparsity level around the range of $5\%-10\%$ is considered in our experiments based on the scenario and the vehicle. Needless to say, parameter sweep analysis is conducted separately for each data-set and each scenario. It is worth mentioning that, for a given $\beta$, the ratio $\lambda/\mu$ can be automatically selected given the desired sparsity range following our method proposed in \cite{murtada_efficient_2021}. However, since for both CADMM and SADMM an empirical search for $\beta$ is needed and at almost exact sparsity levels, a parameter sweep would facilitate finding more accurate parameters for the sake of comparison.
An example of image sparsity and image entropy versus different parameters is shown for the first scenario. For later experiments, such analysis will be omitted for brevity. CADMM and SADMM are run for a maximum number of $100$ iterations while the feasibility tolerances $\epsilon_{abs}$ and $\epsilon_{rel}$ are both set to $10^{-2}$.
Scenario-specific parameters and reconstructed images of all the aforementioned experiments are provided and discussed in the sequel.
\\
\subsubsection{Full Views - Full Bandwidth (FVFB)}
Using the full $360 \degree$ azimuth extent and the full bandwidth of the data-set (approximately equal to $5.35$ GHz), we reconstruct the images of the two vehicles to validate our proposed methods and show their superior reconstruction quality with respect to the conventional BP method. The reconstructed images in this experiment can also be used as a reference for the subsequent scenarios when images are reconstructed using limited views and/or limited bandwidth measurements.
As mentioned previously, to avoid anisotropic scattering over the large angular azimuth extent, it is divided into sub-apertures (clusters) in which point isotropic scattering assumption is valid. The angular width of a cluster has a double-faced effect on the imaging performance trading-off between angular resolution and homogeneous targets scattering. The wider the cluster the better the resolution while the scattering becomes anisotropic. Therefore, in this experiment, the cluster width is chosen to maximize the imaging performance in a balanced manner. We use a cluster width of $5\degree$ which is a good trade-off since it allows for a cross range resolution approximately equivalent to the range resolution provided by the bandwidth while maintaining a minimum degree of scattering anisotropicity of the targets.\par
As anticipated earlier, we perform a hyperparameter sweep over different values of $\beta$ and $\lambda /\mu$ to obtain the values that provide good reconstruction quality given image sparsity level. The sparsity level is decided based on the assumed dimensions of the observed targets relative to the scene dimensions. They are set separately for each scenario due to the dissimilarities of aperture size and signal bandwidth utilized, hence images exhibit a different resolution in each scenario. Fig.~\ref{fig:FVFB_ParSw} shows the normalized sparsity level and entropy of the reconstructed image of the 'Jeep99' data-set versus the varying parameters of $\beta$ and $\lambda/\mu$ for both SADMM and CADMM. The desired sparsity level range is highlighted with a light green color in the figure. As expected, the higher the ratio $\lambda/\mu$, the more sparse are the reconstructed images for both the algorithms and naturally the lower the entropy. This is true except for some values of $\beta$ for which SADMM optimization does not converge to a sparse solution. Such divergence for those parameters can be seen from the entropy values which indicate reconstruction of random images. Moreover, beyond those values of $\beta$, reconstructed images using the same $\lambda/\mu$ ratio exhibit similar sparsity levels and attain close values for entropy as shown in the magnified part of Fig.~\ref{fig:FVFB_ParSw}.\par
\begin{figure}[h!]
	\centering
	\includegraphics[width=3.7 in]{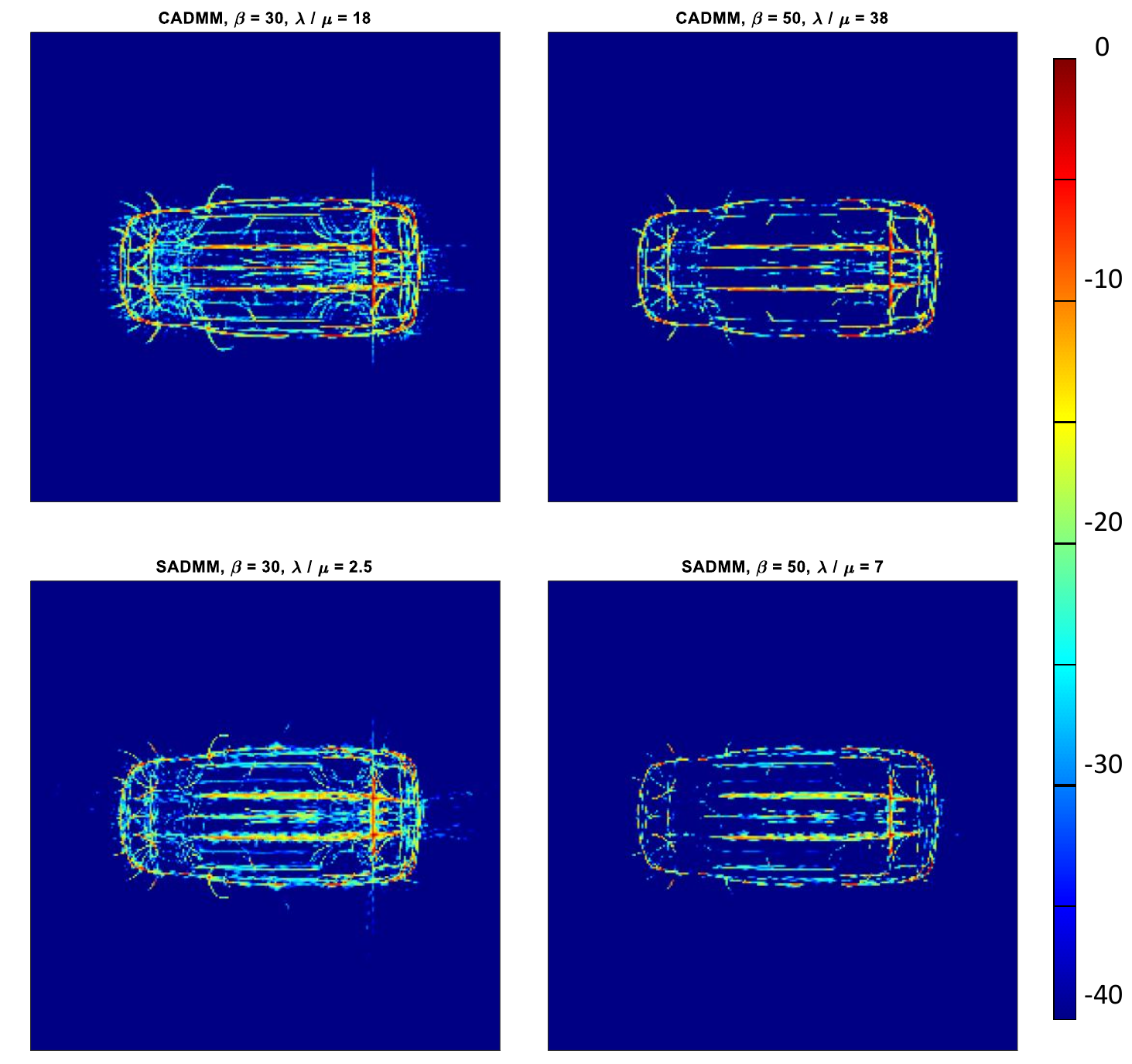}
	\caption{Jeep image reconstruction at different sparsity levels; $12\%$ (left column), $5\%$ (right column)}
	\label{fig:FVLB_ParComp}
\end{figure}
\begin{figure}[ht!]
	\centering
	\includegraphics[width=3.6 in]{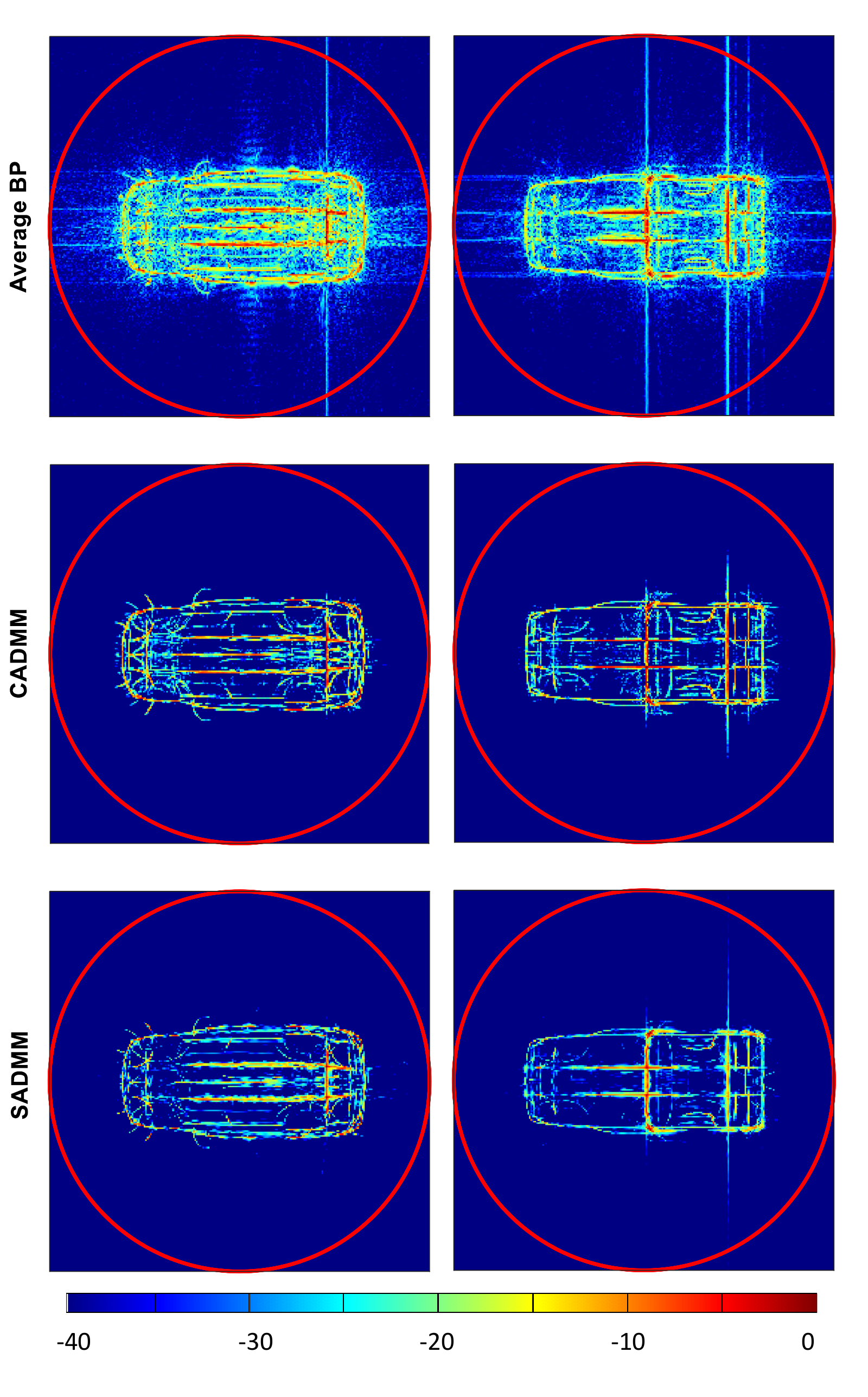}
	\caption{FVFB image reconstruction at sparsity level $\approx 10\%$, $Q=72$ clusters, bandwidth $=5.35$ GHz; Jeep Cherokee (left column), Toyota Tacoma (right column)}
	\label{fig:FVFB}
\end{figure}

To first show the performance of both methods, CADMM and SADMM images reconstructed considering two different sparsity levels are shown for the 'Jeep99' data-set in Fig.~\ref{fig:FVLB_ParComp}. The left column images have a normalized sparsity degree of $12\%$ versus $5\%$ for the images in the right column. For lower sparsity images, CADMM show a sharp high intensity reconstruction of the strong features of the vehicle such as edges and ceiling structure, and a lower intensity reconstruction of the weaker features such as the projection of the tire wheels. On the other hand, SADMM images manifest an averaged intensity of the different parts of the vehicle resulting in less sharp images. While the behavior is maintained for the images at higher sparsity, the weaker features are further suppressed in the images of both methods. Moreover, the images of both methods at a similar sparsity level have similar entropy values.
This behavior is again confirmed by the reconstructed images of 'Tocoma' vehicle. At a sparsity level of $10\%$, the reconstructed images of the two vehicles are shown in Fig~.\ref{fig:FVFB} in addition to the images reconstructed through conventional back-projection averaged over all clusters. The red dots on the images show the angular aperture views which cover $360 \degree$ in this experiment. Note that due to the abundance of the available bandwidth and the full aperture measurements, both CADMM and SADMM images have very high resolution and show the detailed structure of the imaged objects.\par
Additionally, the processing time until termination in SADMM is slightly higher than CADMM. The numerical values of the parameters used in image reconstruction and the corresponding metrics are reported in Table \ref{tab:summary} while the ratio between SADMM and CADMM processing time and number of iterations are reported in Table \ref{tab:proc_time}.\\
%
\subsubsection{Full Views - Limited Bandwidth (FVLB)}
\begin{figure}[ht!]
	\centering
	\includegraphics[width=3.6 in]{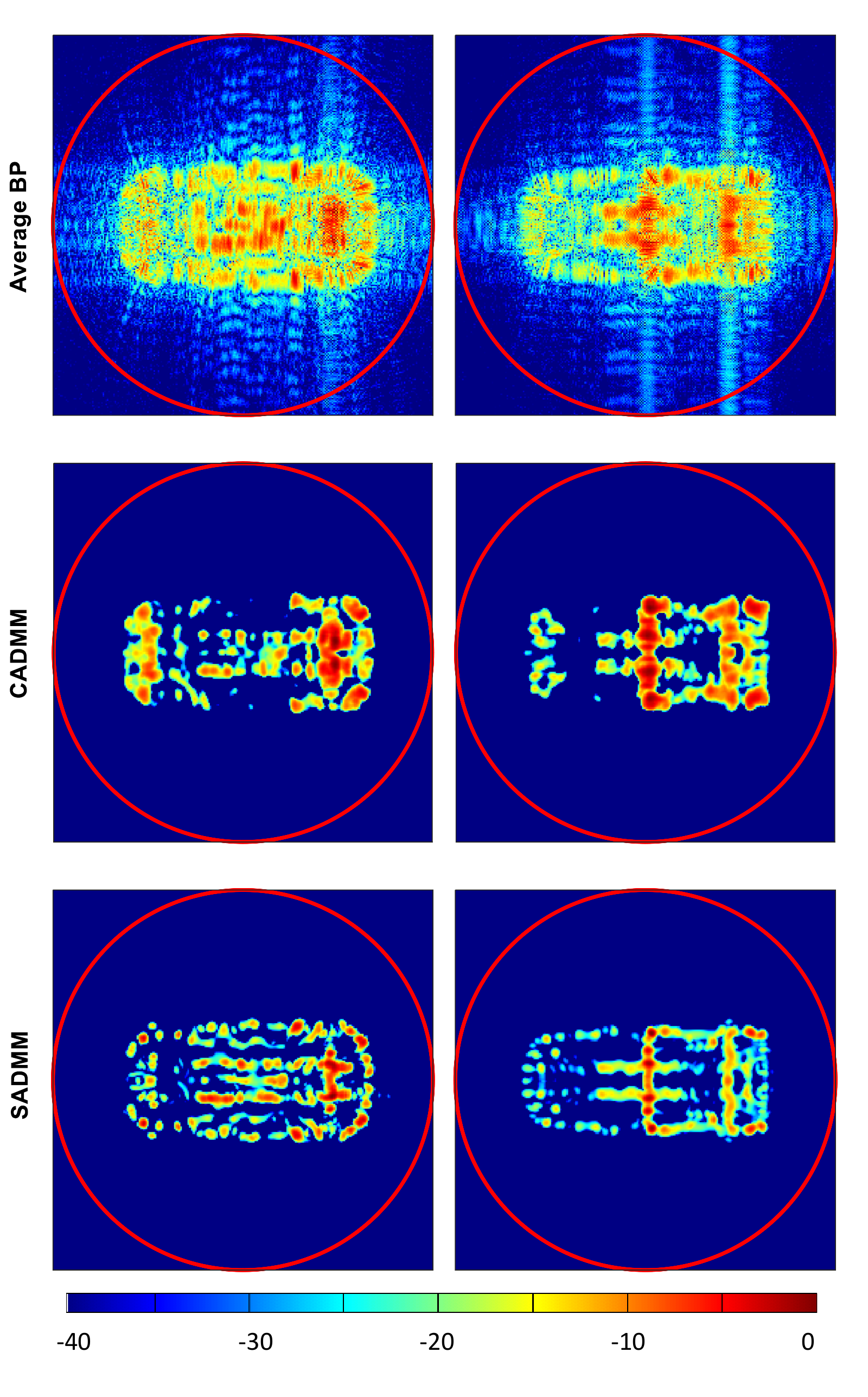}
	\caption{FVLB image reconstruction at sparsity level $\approx 10\%$, $Q=72$ clusters, bandwidth $=600$ MHz; Jeep Cherokee (left column), Toyota Tacoma (right column)}
	\label{fig:FVLB}
\end{figure}
\begin{figure}[h!]
	\centering
	\includegraphics[width=3.6 in]{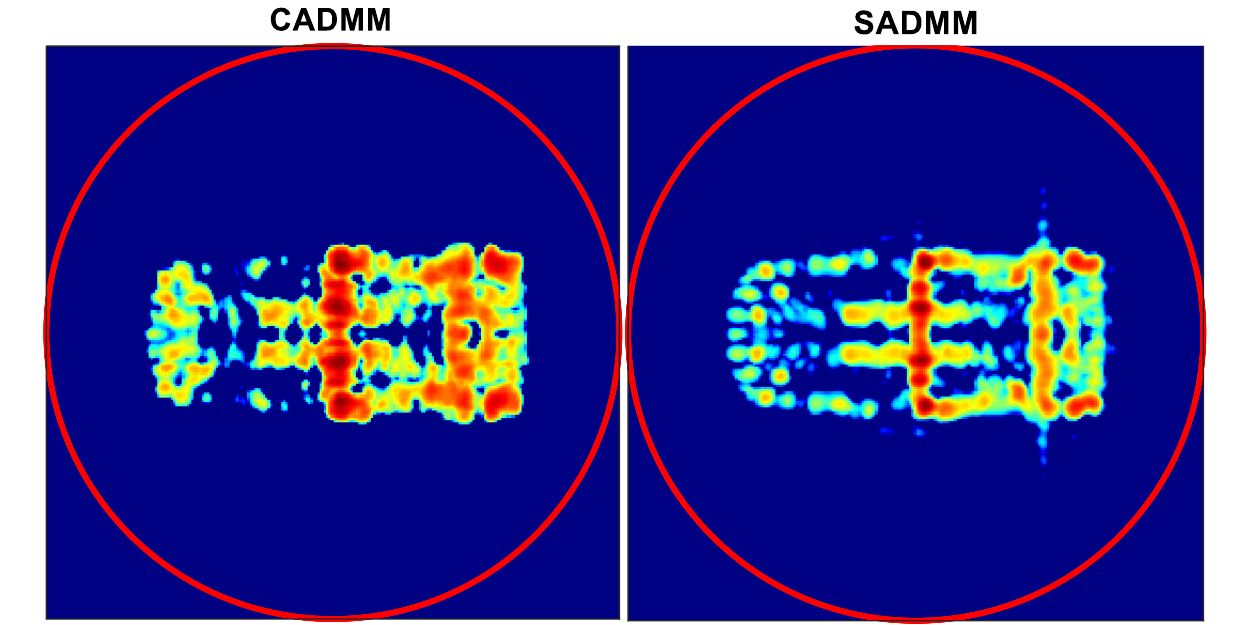}
	\caption{Less sparse FVLB reconstruction of Toyota Tacoma (sparsity level $\approx 15\%$).}
	\label{fig:FVLB_T_E}
\end{figure}
In practice, the signal bandwidth of a SAR system is usually an order of magnitude less than the available bandwidth of the data-set. Thus, analyzing the performance of our proposed algorithms in this case is of high interest and is a relevant use case in WSAR imaging. To realize a limited bandwidth measurement scenario, we utilized an equivalent of $600$ MHz samples around the center frequency from the samples of phase history. In this experiment as well, SADMM images exhibit the property of higher averaging than that of concentrated intensity resulting from CADMM images as shown in Fig.~\ref{fig:FVLB}. This peculiarity makes it capable of capturing the true structure of the vehicles even with low bandwidth measurements when referring to the images in Fig.~\ref{fig:FVFB} of FVFB experiment. On the other hand, images reconstructed using CADMM have higher intensity around the strong reflectors and weaker or no intensity of the poorly reflective components of the vehicles; this is evident in the reconstructed images of both vehicles in Fig.~\ref{fig:FVLB}. For example, the crossing of the beams in the rear part of 'Jeep99' data-set is localized and well identified with strong intensity in SADMM image, while in CADMM image, the same area exhibits only a strong intensity in a wider region. Similarly, in 'Tacoma' images, CADMM fails to capture the outline of the vehicle at this sparsity level and the image is dominated by the strong trunk while in SADMM image the vehicle outline makes an appearance and even the trunk is better identified. The depicted images have a sparsity level approximately equals to $10\%$. Note that by considering lower sparsity, images of both algorithms will have an increase of the background intensity around the strong features without capturing the general structure of the object differently. An example is shown in Fig.~\ref{fig:FVLB_T_E} where the images of 'Tacoma' vehicle are reconstructed at lower normalized sparsity level of about $15\%$. In summary, although CADMM images have similar entropy values as their SADMM counterparts at the same sparsity level, the latter provides higher capability of capturing the structure of the observed targets given relatively low bandwidth measurements. The exact values of sparsity and entropy for each image are reported in Table \ref{tab:summary}.\par
The superior performance of SADMM comes at the cost of increased computational complexity to reach convergence. This complexity is manifested through the higher number of iterations and the longer processing time required for SADMM to reconstruct the shown images. The ratio of these two quantities for SADMM and CADMM is depicted in Table \ref{tab:proc_time}.\\

\subsubsection{Limited views - Limited Bandwidth (LVLB)}
\begin{figure}[h!]
	\centering
	\includegraphics[width=3.6 in]{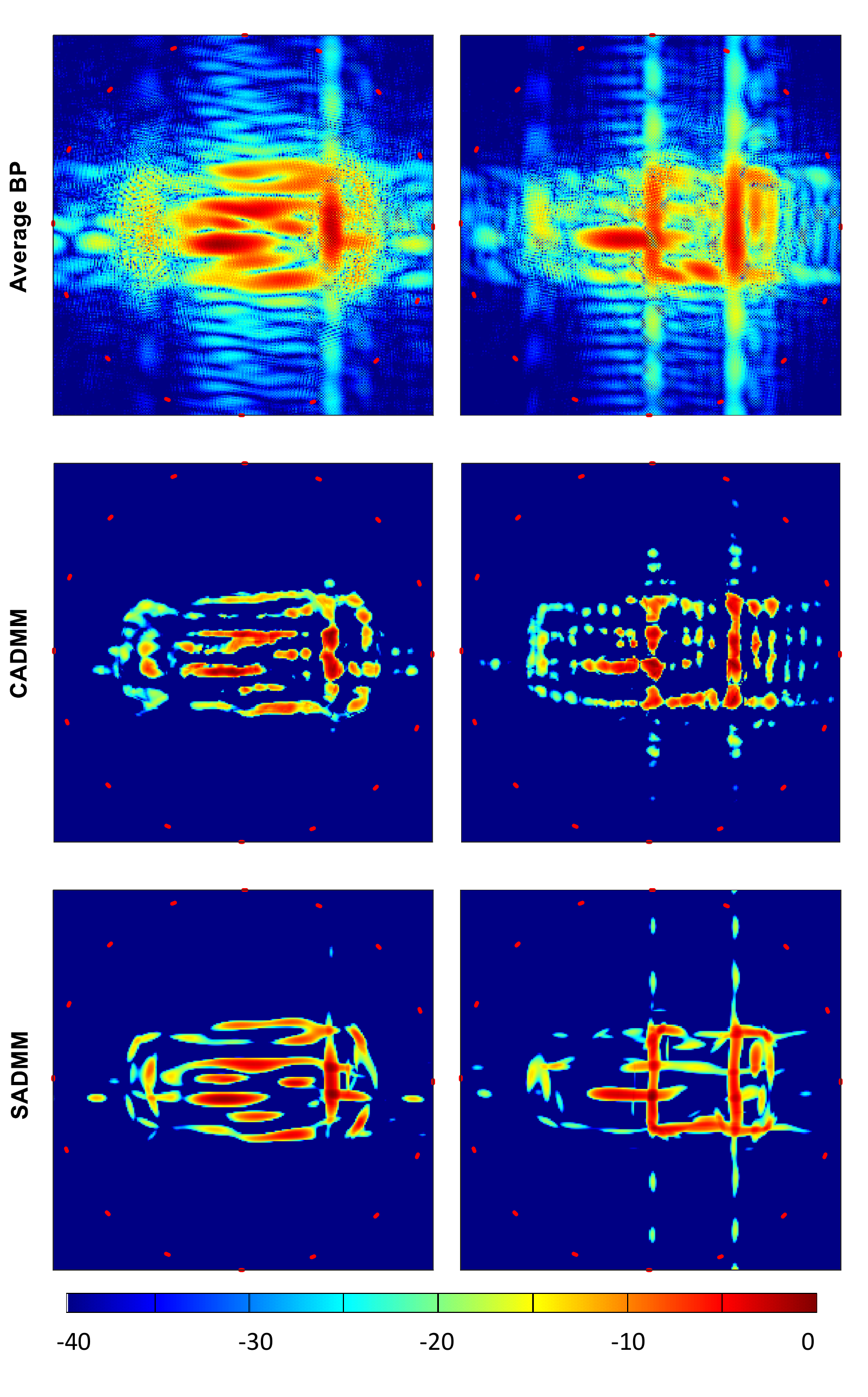}
	\caption{LVLB image reconstruction at sparsity level $\approx 10\%$, $Q=16$ clusters, bandwidth $=600$ MHz; Jeep Cherokee (left column), Toyota Tacoma (right column)}
	\label{fig:LVLB_O0}
\end{figure}
\begin{figure}[h!]
	\centering
	\includegraphics[width=3.4 in]{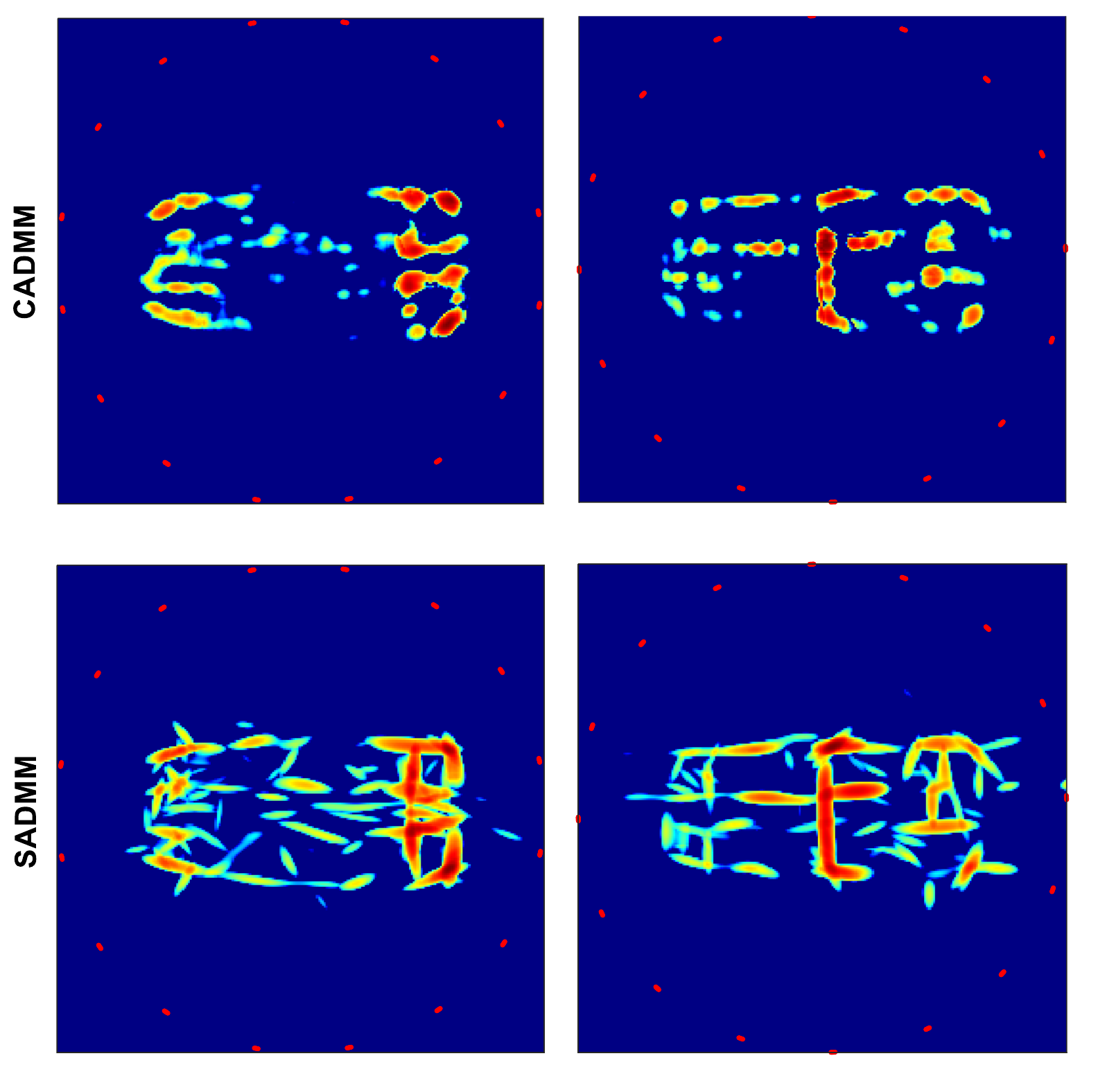}
	\caption{LVLB reconstruction with different random sensors orientation; Jeep Cherokee (left column), Toyota Tacoma (right column)}
	\label{fig:LVLB_O1}
\end{figure}
A mono-static distributed sensing scenario can be realized by considering far-field illumination with limited-narrow views of the full aperture measurements. In this experiment, we consider using the data-set to realize a system of distributed radar sensors in which TDM scheme is used to illuminate the scene of interest where a single cluster transmits at a time. Consequently, in addition to the limited bandwidth measurements of $600$ MHz introduced in the previous experiment, we further consider limited aperture measurements representing the views of distributed sensors. In particular, $16$ clusters of $1 \degree$ width each are uniformly distributed around the scene and considered the viewing angles of the sensors. This number of clusters makes a realistic choice for the number of sensors where they cover only a span of about $4.4\%$ of the full aperture measurements.\par
For this experiment, the reconstructed images of the two vehicles are shown in Fig.~\ref{fig:LVLB_O0}. Similar to the previous experiment, SADMM captures both vehicles' structure better than CADMM. However, due to the limited aperture measurements, the artifacts present in the images of both methods are stronger. Increasing the sparsity would eliminate the artifacts but further limits the reconstruction of the entire outline of the vehicles. Of course, reconstructed images are views dependent. However, the performance of both methods is the same when different orientation of the sensors is considered. For example, the images reconstructed using another random orientation of views are shown in Fig.~\ref{fig:LVLB_O1}. The images confirm the capability of SADMM to retain the original structure of the imaged target while CADMM has a better ability to diminish the artifacts. On another note, processing times of both algorithms are roughly similar in this experiment given a limited amount of measurements. Finally, the parameters used to reconstruct the images in Fig.~\ref{fig:LVLB_O0} and the corresponding values of entropy and sparsity are reported in Table \ref{tab:summary} while processing time and the number of iterations ratios are reported in Table \ref{tab:proc_time}.\par

\begin{table*}[h!]
\centering
\caption{Summary of parameters used in the experiments and corresponding metrics}
\label{tab:summary}
\begin{tabular}{@{}clllllll@{}}
\cmidrule(l){3-8}
\multicolumn{1}{l}{} &
   &
  \multicolumn{2}{c}{FVFB} &
  \multicolumn{2}{c}{FVLB} &
  \multicolumn{2}{c}{LVLB} \\ \cmidrule(l){3-8} 
\multicolumn{1}{l}{} &
   &
  \multicolumn{1}{c}{CADMM} &
  \multicolumn{1}{c}{SADMM} &
  \multicolumn{1}{c}{CADMM} &
  \multicolumn{1}{c}{SADMM} &
  \multicolumn{1}{c}{CADMM} &
  \multicolumn{1}{c}{SADMM} \\ \midrule
\multirow{2}{*}{($\beta$, $\lambda/\mu$)} & Jeep   & (30,18)  & (30,2.5) & (30,16) & (30,1.4) & (10,1.25) & (20,0.4)  \\ \cmidrule(l){2-8} 
                                          & Tacoma &  (30,25) & (30,2.5) & (30,24)  & (50,1.8)  & (10,1.75) & (20,0.65) \\ \midrule
\multirow{2}{*}{Sparsity}                 & Jeep   & 0.100 & 0.102  &  0.104 & 0.102 & 0.100  & 0.098 \\ \cmidrule(l){2-8} 
                                          & Tacoma &  0.096 & 0.10 & 0.097  & 0.098  & 0.099 & 0.103  \\ \midrule
\multirow{2}{*}{Entropy}                  & Jeep   &  1.24 & 1.25 & 1.27 & 1.26 & 1.25  & 1.22 \\ \cmidrule(l){2-8} 
                                          & Tacoma &  1.19 & 1.24 & 1.20 & 1.21 & 1.24 & 1.26 \\ \bottomrule
\end{tabular}
\end{table*}

\begin{table*}[!h]
\centering
\caption{Relative convergence and complexity}
\label{tab:proc_time}
\begin{tabular}{@{}ccccccc@{}}
\toprule
       & \multicolumn{2}{c}{FVFB} & \multicolumn{2}{c}{FVLB} & \multicolumn{2}{c}{LVLB} \\ \midrule
    \begin{tabular}[c]{@{}c@{}} Ratio\\SADMM/CADMM\end{tabular}&
  \begin{tabular}[c]{@{}c@{}}Number of\\  iterations \end{tabular} &
  \begin{tabular}[c]{@{}c@{}}Processing\\ time \end{tabular} &
  \begin{tabular}[c]{@{}c@{}}Number of\\  iterations \end{tabular} &
  \begin{tabular}[c]{@{}c@{}}Processing\\ time \end{tabular} &
  \begin{tabular}[c]{@{}c@{}}Number of\\  iterations \end{tabular} &
  \begin{tabular}[c]{@{}c@{}}Processing\\ time \end{tabular} \\ \midrule
Jeep   &     1.41    &    1.19     &      3.45   &    2.11      &     1.45    &     1.20    \\ \midrule
Tacoma &     1.41    &    1.22     &      3.13   &    2.06      &     1.51    &     1.04    \\ \bottomrule
\end{tabular}
\end{table*}
To summarize, in the first experiment where a plenitude of measurements in both aperture and bandwidth is available, both CADMM and SADMM are capable of reconstructing detailed and super-resolution images of the observed targets far surpassing the conventional methods. On the other hand, in the latter experiments where measurements are limited in aperture and/or bandwidth, SADMM exhibits superior performance over CADMM in terms of capturing the structure of the target and reconstructing smoother images. Although they have similar entropy in all cases, the depicted images reconstructed by both algorithms show a clear visual advantage of SADMM when compared with the images of full measurements. Such higher quality comes at the expense of computational cost. Surprisingly, in terms of convergence and complexity, SADMM fell behind the most in the second experiment where the full aperture measurements with limited bandwidth are considered. This can be owed to the fact that CADMM features a high convergence rate requiring less than a third of SADMM iterations to reach such concentrated intensity images and by having less degrees of freedom. 
\section{Conclusion}
\label{sec:conc}
In this paper, a novel approach for widely distributed radar imaging based on ADMM framework is proposed. Sparsity prior has been imposed on a defined global image assumed to represent an aggregate view of the scene. Then, developing on top of our previous work, the problem formulation is tailored to this approach and a new formulation has been introduced. The two formulations named CADMM and SADMM are designed to mathematically stipulate the relationship between the images of individual sensors and the global image. The solutions to the proposed formulations have been provided as iterative algorithms that are flexible and amenable to be implemented on a distributed architecture. We have demonstrated the performance of our proposed algorithms through several experiments and showed their significant edge over conventional methods in terms of reconstructed images quality. Moreover, we showed that SADMM outperforms CADMM by reconstructing images of high resolution that better exhibit the structure and the shape of the observed objects, especially when the measurements are limited in bandwidth and/or sparse in aperture. As we illustrated in our experiments, our proposed algorithms are applicable in many scenarios of distributed radar systems and WSAR imaging. Following our approach, various formulations can be further studied and developed either by imposing different prior on the global image and/or by imposing alternative associations with the images of the individual sensors.
\ifCLASSOPTIONcaptionsoff
  \newpage
\fi

\bibliographystyle{IEEEtran}
\bibliography{bibliography}

\end{document}